# Unsupervised dimensionality reduction of polarimetric data for pixel-wise pathological tissue differentiation

**Mickaël Li,[a] Nan Zeng,[a] Liangyu Deng,[a] Mingzhou Jiang,[a] Chang Wu,[b] and Honghui He[a,*]**

[a]Guangdong Research Center of Polarization Imaging and Measurement Engineering Technology, Shenzhen Key Laboratory for Minimal Invasive Medical Technologies, Institute of Biopharmaceutical and Health Engineering, Tsinghua Shenzhen International Graduate School, Tsinghua University, Shenzhen 518055, China

[b]Shenzhen Sixth People's Hospital (Nanshan Hospital) Huazhong University of Science and Technology Union Shenzhen Hospital, Shenzhen 518052, China

**Abstract**. Extracellular matrix (ECM) constitutes a key basement structure to human organisms by acting as a complex network of large proteins and carbohydrates that provide structural support to surrounding cells. Remodeling in the extracellular matrix's structural fibers leads to insight into the development of diseases such as cancer, fibrosis and carcinoma. While standard tissues visualization in the ECM involves multiple lengthy histopathological staining protocols, Mueller matrix-based polarimetry provides label-free tissue slices' microstructural information and optical properties. This work aims to identify three types of fiber tissues commonly found in the ECM of gastrointestinal tissue specimens by analyzing their polarization properties. To address decomposition methods' reliance on restrictive hypotheses and inability with an individual polarization-based parameter to determine the nature of a given biological tissue; this study employs Uniform Manifold Approximation and Projection (UMAP) method to offer greater discriminative power and flexibility. Subsequently, polarization-based features will be extracted and compared between fiber regions statistically to discern potential diagnostic differences. By providing colorized images, this work aims to demonstrate the feasibility of distinguishing different fibers with polarization approach, offering insights for future clinical development while complementing existing staining methods for pathological tissue specimens.

*He Honghui, E-mail: he.honghui@sz.tsinghua.edu.cn

## 1 Introduction

The extracellular matrix (ECM) serves as a structural scaffold for microorganisms, providing mechanical support while facilitating cell adhesion, migration, and signaling. The ECM is primarily composed of proteins such as collagen (with type I dominant in collagen fibers and type III in reticular fibers), elastin (in elastic fibers), glycosaminoglycans, proteoglycans, and enzymes, and therefore regulates tissue development while acting as a reservoir for bioactive molecules. These biochemical components each play distinct roles in maintaining tissue integrity and function: collagen fibers, abundant in tendons, ligaments, skin, and other fibrous tissues provide tensile strength and stretch resistance to these structures; reticular fibers (type III collagen coated with glycoproteins and proteoglycans) form supportive networks in soft organs and lymphoid stroma; and elastic fibers enable stretch-recoil properties in skin, lungs, and blood vessels. By varying in composition across tissues, the ECM ensures both structural integrity and functional adaptability. These three types of fiber structures provide critical insights into many diseases. Modifications in their physical properties such as loss or degradation or proliferation directly relate to pathological stage progression[1-3]. Each type of fiber corresponds to distinct pathological conditions, as they influence different tissue regions and exhibit varying physical properties and functionalities. To highlight these issues for clinicians, traditional histopathology relies on staining processes applied to continuous thin tissue slices, enabling visualization of different tissue structures. Masson's trichrome (MT), Gomori's silver (G&S), and Verhoeff's Van Gieson (EVG) stains are commonly used to detect collagen fibers, reticular fibers and elastic fibers. While histological staining remains the indispensable gold standard for determining chemical specificity, the multi-step protocols required to differentiate multiple fiber types can be labor-intensive and susceptible to morphological discrepancies between adjacent sections. In this context, Mueller matrix microscopy combined with modern analysis techniques serves as a valuable complementary tool. It offers a rapid, label-free dimension of structural information in a single scan, enriching the analysis without the complexities of sequential processing.

In the past decades, Mueller matrix-based polarimetry has emerged as a promising tool for enhancing contrast in tissue structures, detecting early disease biomarkers, and improving diagnostic accuracy in biomedical imaging. As a non-destructive, label-free imaging technique, it has gained increasing popularity and field of applications. In a clinical setting, polarized light provides a wealth of microstructural information about biological samples[4-9], enabling non-invasive characterization of fibrous structures such as collagen and its interactions with cancerous tumor[10,11], fibrosis[12], and other pathologies[13,14]. Recent advancements in polarimetry techniques focus on extracting deeper polarization insight[15] from specimens using modern technology including artificial intelligence and advanced simulations. Furthermore, novel approaches involving vectorial light fields and adaptive optics are being explored to ensure imaging robustness against complex aberrations[16], demonstrating the potential of this technology in future medical diagnostics. Since a single Mueller matrix parameter is insufficient to fully characterize the nature of a tissue, this work seeks to overcome the limitations of traditional Mueller matrix analysis by synergistically integrating insights from multiple calculation methods. This approach aligns with multi-parametric fusion techniques developed to capture complete structural information previously proposed. For instance, Zhai et al. combined multiple parameters to better characterize fibrous tissues[17], while Roa et al. discriminated between collagen and elastic fibers by compiling Mueller matrix components and decompositions into individual information channels to form a 3D voxel for classification and analysis[18]. In this study, to demonstrate an unsupervised dimensionality reduction of polarimetric data for pixel-wise pathological tissue differentiation, we focus on the gastrointestinal (GI) tract's ECM, which plays a critical role in colorectal cancer (CRC)—the third most diagnosed cancer worldwide. CRC accounts for over 3 million deaths annually, representing nearly 10% of all global cancer cases and fatalities. A key pathological feature of CRC and other cancer pathology is the disruption of the ECM's fibrous structure, contributing to disease progression[19,20]. Through the strategic combination of complementary polarization parameters within a dimension-reduction framework, we enhance analytical applicability, reduce system ambiguity, and improve accuracy of core histopathological properties to facilitate clinical visualization and spatial localization.

## 2 Materials and Methods

*2.1 Colon tissue samples preparation*

A total of 9 human colon 4 μm thick tissue slices were provided by Shenzhen Sixth People's Hospital (Nanshan Hospital) Huazhong University of Science and Technology Union Shenzhen Hospital. Each sample was fixed in formalin and embedded in paraffin to be sectioned in four transversely. Then, sections were deparaffinized with xylene and mounted on glass slides and dried naturally. Due to the limitations of H&E staining in visualizing certain fiber types, specialized staining techniques are often required. In this study, GI adjacent serial sections were stained with H&E, MT, G&S and EVG dyes to delineate specific fiber types. This study was approved by the Ethics Committee of Tsinghua Shenzhen International Graduate School, Tsinghua University.

*2.2 Mueller matrix derived parameters*

A Mueller matrix (MM) is a 4x4 real-valued matrix that represents modulation from media on polarization state of light upon their interaction as described in Eq. (1). Although the physical meanings of each Mueller matrix element remain challenging to interpret, various decomposition methods[21-24] have introduced a set of more intuitive polarization-based parameters from the original MM. However, each decomposition method operates with specific theoretical hypothesis and constraints, restricting most practical studies to focus under single decomposition framework or compare limited approaches on selected samples. For instance, Lu-Chipman's polar decomposition, or the Mueller matrix polar decomposition (MMPD)[21] constitutes a product decomposition-based method assuming incidence light's polarization transformation occurring in series. In biological samples, the most adopted MMPD form assumes that polarization state transformation happens in the order of diattenuation, retardation and depolarization, suggesting the decomposition of an arbitrary MM into the product of three elementary matrices illustrated as Eq. (2). Since then, physical quantities such as diattenuation, depolarization, and retardance can be expressed with these elementary matrices shown as Eqs. (3)-(5).

$$M = \begin{pmatrix} M_{11} & M_{12} & M_{13} & M_{14} \\ M_{21} & M_{22} & M_{13} & M_{24} \\ M_{31} & M_{32} & M_{33} & M_{34} \\ M_{41} & M_{42} & M_{43} & M_{44} \end{pmatrix}, \tag{1}$$

$$M = M_\Delta M_R M_D \tag{2}$$

$$D = \frac{1}{m_{11}}\sqrt{m_{12}^2 + m_{13}^2 + m_{14}^2}, \tag{3}$$

$$\Delta = 1 - \frac{1}{3}|\text{tr}(M_\Delta) - 1|, \tag{4}$$

$$R = \cos^{-1}\left(\frac{1}{2}tr(M_R) - 1\right), \tag{5}$$

In addition to the MMPD, the differential decomposition illustrated as Eq.(6) provides optical parameters of a sample under different hypotheses, enabling the retrieval of $L_m$ and $L_u$ shown as Eq.(7), which respectively encompasses all nondepolarizing transformations (linear retardation in Eq.(8), linear diattenuation in Eq.(9)) and contains purely depolarizing properties ($a_{22}$, $a_{33}$ in Eq.(10)) with G the Minkowski metric tensor[22].

$$\frac{dM}{dz} = mM, \ln M = mz = L_u + L_m \tag{6}$$

$$L_m = \frac{1}{2}(L - GL^T G), L_u = \frac{1}{2}(L + GL^T G) \tag{7}$$

$$\delta_L = \sqrt{L_m(2,4)^2 + L_m(3,4)^2} \tag{8}$$

$$D_L = \sqrt{L_m(1,2)^2 + L_m(1,3)^2} \tag{9}$$

$$a_{22} = L_u(2,2), a_{33} = L_u(3,3) \tag{10}$$

Similarly, the Mueller matrix transformation (MMT) method derives a set of polarization-based parameters A, b, t (Eqs. (11)-(13)) closely associated to a sample's microstructural features as they are obtained straight from the MM elements without relying on matrix decomposition[23]. Other popular decomposition techniques also exist, such as Cloude decomposition, which enables the analysis of polarization purity and is particularly well-suited for studying depolarizing media and their depolarization properties[24,25]. However, all these methods' preliminary requirements may not always be held, the MMPD assumes a specific order of light-path encounters; some of the MMT derived parameters exhibit instability; the differential decomposition relies on a positive determinant and assumes longitudinal homogeneity across samples. The potential for such misinterpretation of the model, given the actual structure of the sample, highlights the need for additional, deeper studies[26-30]. While combining the strengths of the above techniques shows promise in tissue polarimetry, systematic integration remains unexplored.

$$A = \frac{2bt}{b^2+t^2} \tag{11}$$

$$b = \frac{1}{2}(m_{22} + m_{33}) \tag{12}$$

$$t = \frac{1}{2}\sqrt{(m_{22} - m_{33})^2 + (m_{23} - m_{32})^2} \tag{13}$$

*2.3 Experimental setup*

Here measurement procedures are conducted on a dual-rotating retarder-based Mueller matrix microscope which consists mainly of a polarization state generator (PSG) and a polarization state analyzer (PSA) module, as shown in Fig.1. Both PSG and PSA modules are composed of a linear polarizer P (extinction ratio 1:5000 450-700 nm, Lbtek Co. Ltd., China) and a quarter waveplate R (633 nm, Lbtek Co. Ltd., China) set on a motorized rotatable stage (DDR25/M, Thorlabs Inc., USA) with bidirectional repeatability up to 60 μrad. The L1 and L2 (Hengyang Optic, China) are lenses ensuring respectively collimation of incident light (LED 633 nm, Hengyang Optics, China) for rays' parallelism and collection of emergent light on to the focal plane of a grayscale CMOS camera (1080*1440, MV-CA016-10UM, Hikvision, China), raw image data are acquired under a 10x magnification (10x 0.25 NA, 400-700 nm, Olympus, Japan). It should be noted that the current method is optimized for this specific resolution and that significantly altering the objective (e.g., lowering to 4x) would increase the pixel-averaging effect and potentially impact the analysis results. Future clinical implementations must standardize the optical magnification to ensure consistent diagnostic thresholds. The MM microscope is calibrated according to the eigenvalue calibration method[31], assuming an experimental error below 1%. In each measurement, the quarter waveplates R1 and R2 rotate at different angular speeds $\omega_2 = 5\omega_1$ inducing a modulation of polarization state, therefore, MM elements are obtained with Fourier analysis[32].

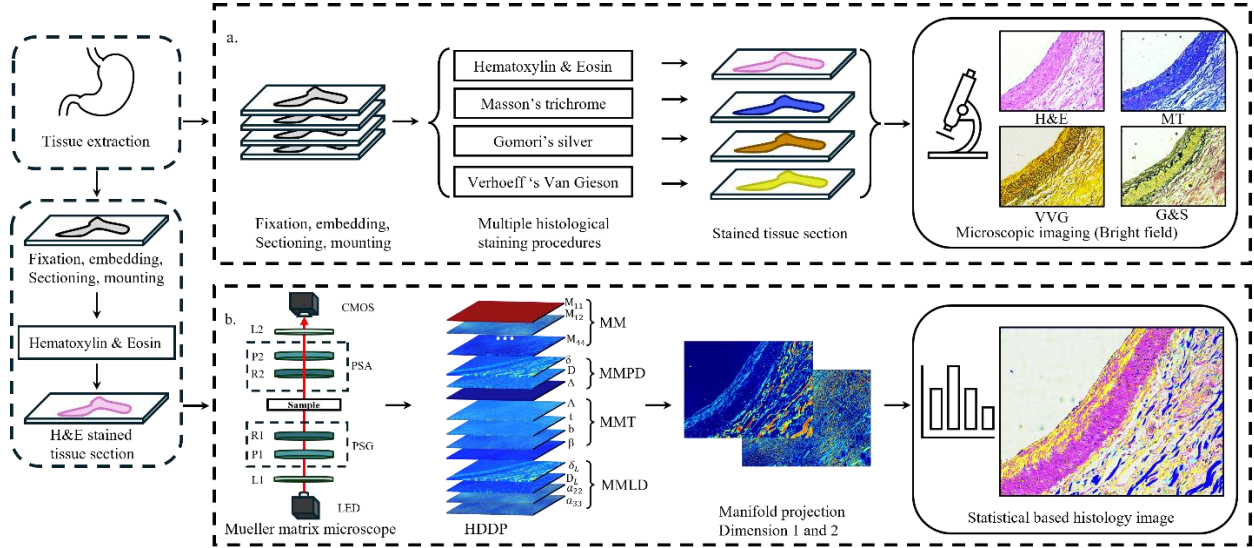

**Fig 1** Schematic comparison of the conventional and polarimetric histopathology workflows. (a) Conventional characterization of fiber types relies on adjacent serial sections stained with specific dyes (e.g., Masson's Trichrome, Gomori's Silver, Verhoeff-Van Gieson). While providing chemical specificity, this process requires multiple slides and labor-intensive protocols. (b) The proposed polarimetric method is performed directly on a single H&E-stained section (which also provides morphological context). The sample is imaged using a Mueller matrix microscope (P: polarizer, R: quarter wave plate, L: collimating lens) to generate the Mueller matrix. Subsequent decomposition analysis (e.g., polar, differential decomposition) and statistical dimensionality reduction (UMAP) extract microstructural features, enabling the differentiation of specific tissue regions on the H&E slide without the need for additional specialized staining.

*2.4 Statistical analysis*

Given the inherent complexity of biological tissues that often exceed the idealized theoretical frameworks mentioned above, this work systematically integrates multidimensional MM analytics through a synergistic fusion of complementary decomposition methods (always within their respective validity frameworks) and dimension-reduced parameter spaces using the Uniform Manifold Approximation and Projection method[33]. Mueller matrix imaging was performed directly on H&E-stained sections to ensure morphological correspondence and demonstrate clinical compatibility, generating high-dimensional data directly from standard diagnostic slides with enhanced polarimetric features of the tissue structures[34,35]. These polarization elements form the ambient dimensions of the manifold technique and each HDDP with its full set of dimensional values encrypts sample's local microstructural features, resulting in a total of 1080*1440 HDDP map as illustrated in Fig. 1.

Given the unsupervised nature of the UMAP algorithm, the model prioritizes identifying the intrinsic structures of the data. Therefore, instead of using labels for training, we utilized standard histological analysis ground truth as a critical benchmark for validation. Specifically, an experienced pathologist examined the stained sections to identify and annotate tissue structures. These expert-verified annotations were then used to biologically validate the clusters identified by the optical method. The first step in UMAP involves the identification of the k-nearest neighbors for each HDDP. In high-dimensional spaces, the process becomes exponentially complexified due to the intensively increasing computations required to determine the nearest neighbors in the algorithm for every pixel sample. Additionally, excessively irrelevant or noisy input data distorts the manifold projection, structure on random fluctuations will be imposed by the algorithm, leading to physically meaningless clustering artifacts rather than genuine fiber characteristics, restricting the number of input parameters. Obviously, relevant features can enhance the significance of distance calculations in UMAP. As a result, clustering results are more accurate because the measured "distance" better reflects the actual differences between points in terms of physical features. Since linear birefringence is the dominant polarization property in fibers, most input parameters building the HDDP will be retardance-related. However, to account for structural variability in fibers, additional features representing diattenuation and depolarization are also included in the training process to detect localized inhomogeneities. Therefore, the feature set was constructed using MM elements and MM derived parameters including $\delta$, $D$ and $\Delta$ from polar decomposition; $\delta_L$, $D_L$, $a_{22}$ and $a_{33}$ from differential decomposition; and $A$, $b$, $t$ and $\beta$ from Mueller matrix transformation. Consequently, each pixel is represented by a total of 30 input dimensions. Despite channels' information are seemingly redundant, in this study, they are retained to provide additional insight upon use of dimension reduction method, as different decomposition methods may reveal nuances in the same-said physical parameter. Outliers associated with residues artifact misadded during slices preparation and de-waxing are removed from the input with the median filtering of neighboring pixels. In order to extract fiber characteristics on the manifold projection, one needs to first identify key features on the projected map. Considering the structural diversity of fibers ranging from thin, delicate and branched reticular networks to thick bundled formations in collagen fibers, to preliminary differentiate fiber types, we propose to assess statistical features of the manifold projection. These statistical features explore local structural information hidden in polarimetric imaging, enabling quantitative characterization of microstructural variations among tissue fibers. The following parameters were analyzed: mean for the central tendency; standard deviation (std) for data dispersion around the average; entropy to assess data unpredictability and randomness; kurtosis for tail heaviness and outlier presence; skewness for detecting distribution and gradient of pixel intensity for revealing alignment, edges, density, and dye uniformity. Downsampling can improve computation efficiency, however, minimal downsampling is preferable in this study as it leads to loss of spatial details in the image going against the preceding paragraph. Thin fiber networks and texture patterns in the image become blurred or disappear if their features are smaller than the new pixel spacing after down sampling. The extent of information loss increases with the sampling factor and varies with image complexity, potentially obscuring the subtle inter-parameter differences albeit deriving from identical physical quantity under different decomposition methods.

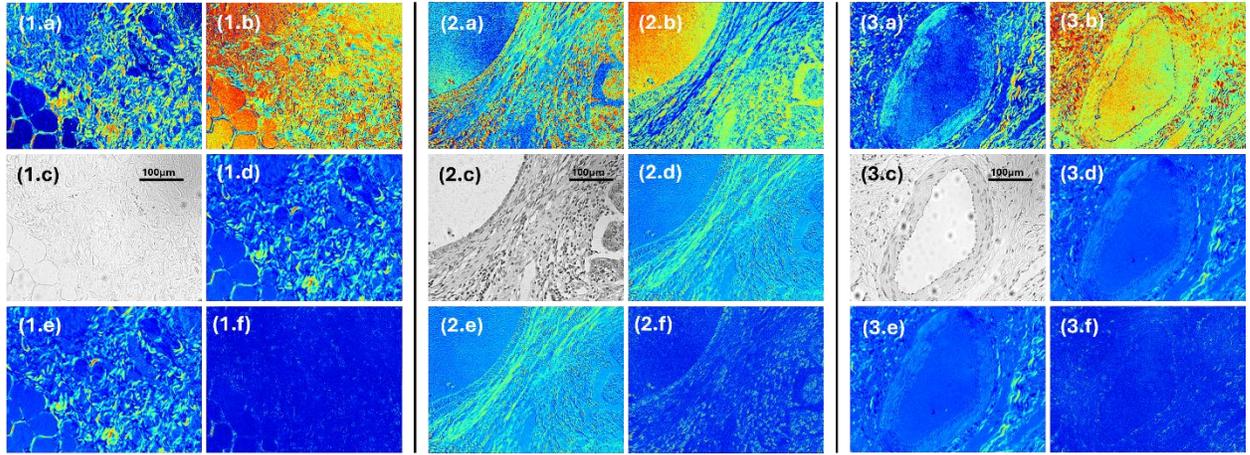

**Fig 2** Tissue fibers polarimetric feature space visualization: collagen (1), reticular (2), and elastic (3) fiber in (a,b) UMAP dimensions, (c) Gray scaled H&E image, (d) Linear retardance (polar decomposition), (e) Linear retardance (differential decomposition), (f) t parameter (Mueller matrix transformation)

## 3  Results

Data acquisition and pre-processing were conducted in accordance with protocol outlined in Section 2 and illustrated in Fig.1. The manifold projections of the HDDP and original grayscale image for distinct tissue fibers are shown in Fig.2 (a–c). The UMAP projection closely resembles the linear retardance obtained from decomposition methods but exhibits a naturally stronger signal and greater clarity and contrast compared with the previously obtained polarization-based parameters in Fig.2 (e-f). To map specific fiber labels from specialized stains to Mueller matrix images, an ROI-based annotation strategy was employed. Anatomically distinct fiber bundles were visually located on the H&E/Mueller matrix images, effectively mitigating registration errors inherent to serial sections. This approach further enhanced computational efficiency by isolating structures prior to manifold projection, thereby reducing the processing load.

For quantitative assessment, the projections were divided into 32 × 32-pixel patches, encompassing 15 distinct ROIs across the tissue samples, the ROIs intersecting with these patches are retained as the cornerstone for statistical calculation. Across all the tissue slices provided, intersection results data corresponding to the same tissue type are grouped together to perform statistical analysis. As the methodology is inherently statistical, the analytical focus is on the quantitative measure of pixel count within each defined ROI. As shown in Figure 3, the projection of histologically guided ROIs reveals well-separated clusters, statistically illustrating the distinct behavioral characteristics of different fibers. The resulting distinct distribution ranges of centralized pixel values confirm the method's ability to differentiate fiber structures based on their intrinsic optical signatures and statistical pixel distributions, effectively distinguishing fiber types without the need for labels.

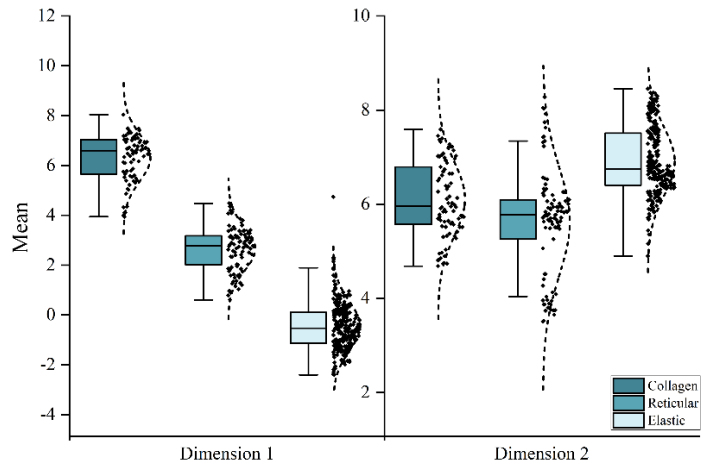

**Fig 3** Boxplot comparison of UMAP embeddings' two-dimensional mean indicator patch distributions, evaluated across different fiber patch datasets

Table I reveals a structured progression in fiber characteristics. Regarding separability, the mean values of the first UMAP component (Dim1) create a clear discriminatory spectrum, positioning elastic fibers in the negative range (-0.47), reticular in the intermediate (2.64), and collagen in the high positive range (6.30). Statistical data reveal significant differences between different fiber categories, highlighting unique distributional tendencies for each fiber type. Regarding textural complexity, a distinct inverse relationship can be observed between entropy magnitude and its variance: from collagen fibers to elastic fibers, entropy increases (from 2.12 to 2.50) while standard deviation decreases (from ±0.38 to ±0.16). Indicating that elastic fibers possess higher information content and texture complexity while exhibiting the highest intra-class consistency. In contrast, collagen fibers feature relatively simpler structures. Regarding distributional shape, collagen stands out with high kurtosis (4.23) and strong negative skewness (-1.37), indicating a heavy-tailed, asymmetric distribution. This means most collagen fiber pixel values cluster in a high-value region, but a long tail extends into the low-value zone. This unique non-Gaussian characteristic makes it an identifiable fingerprint. In contrast, elastic fibers exhibit near-Gaussian characteristics (Skewness ~0.14, Kurtosis ~-0.07). This shift from asymmetric to symmetric distributions indicates fundamental differences in the imaging characteristics of different fiber types and can serve as a robust statistical signature for fiber classification. Collectively, the key statistical metrics derived from the UMAP projection components exhibit unique fiber type characteristics and suggest intrinsic textural differences. Thus, it provides a robust statistical foundation for enhancing the identification and classification of fibers in histopathological images.

Table 1 Statistical metric data summary

| | | Mean | Inter-patch std | Entropy | Kurtosis | Skewness | Gradient |
|---|---|---|---|---|---|---|---|
| Collagen | Dim1 | 6.2984±1.1073 | 1.0565±0.7377 | 2.1262±0.3797 | 4.2328±9.7507 | -1.3661±1.2568 | 0.3710±0.2734 |
| | Dim2 | 4.7486±1.4897 | 0.7389±0.5113 | 2.2197±0.3385 | 3.0545±9.3583 | -0.0440±1.5361 | 0.2704±0.2301 |
| Elastic | Dim1 | -0.4709±1.0907 | 1.9873±0.6883 | 2.4961±0.1566 | -0.0728±1.5265 | 0.1390±0.7251 | 0.6881±0.1915 |
| | Dim2 | 6.1752±1.4424 | 1.6901±0.8883 | 2.4693±0.1469 | 0.0590±1.1773 | -0.3122±0.5645 | 0.6166±0.3231 |
| Reticular | Dim1 | 2.6421±1.1272 | 1.5484±0.6155 | 2.3761±0.2490 | 1.0244±2.9045 | -0.4163±0.9139 | 0.5737±0.2394 |
| | Dim2 | 4.0593±2.1420 | 1.7853±0.5669 | 2.4832±0.2284 | 0.1065±2.1139 | 0.3519±0.7491 | 0.6701±0.2744 |

Finally, building on a statistical model of pixel behavior, we present a physics-inspired computational technique for the pseudo-staining of tissue fibers, offering a novel tool for digital pathology. Figure 4 demonstrates the colorization of specific fiber structures directly on an H&E-stained source image with comparison to their respective fiber emphasizing dye. Collagen fibers are dyed blue, reticular fibers are colored green, and elastic fibers appear in yellow. To facilitate comparison, we deliberately selected regions where two fiber types are visible simultaneously, allowing for clear visualization of the colorization results and comparison. Figure 4 serves to illustrate the comprehensive information capacity of the proposed method. For practical clinical translation, we envision decomposing this high-dimensional dataset into simplified, targeted views. By selectively presenting specific parameter data relevant to a given diagnostic task—rather than the simultaneous overlay shown here—pathologists can assess distinct structural features with greater clarity and without visual clutter.

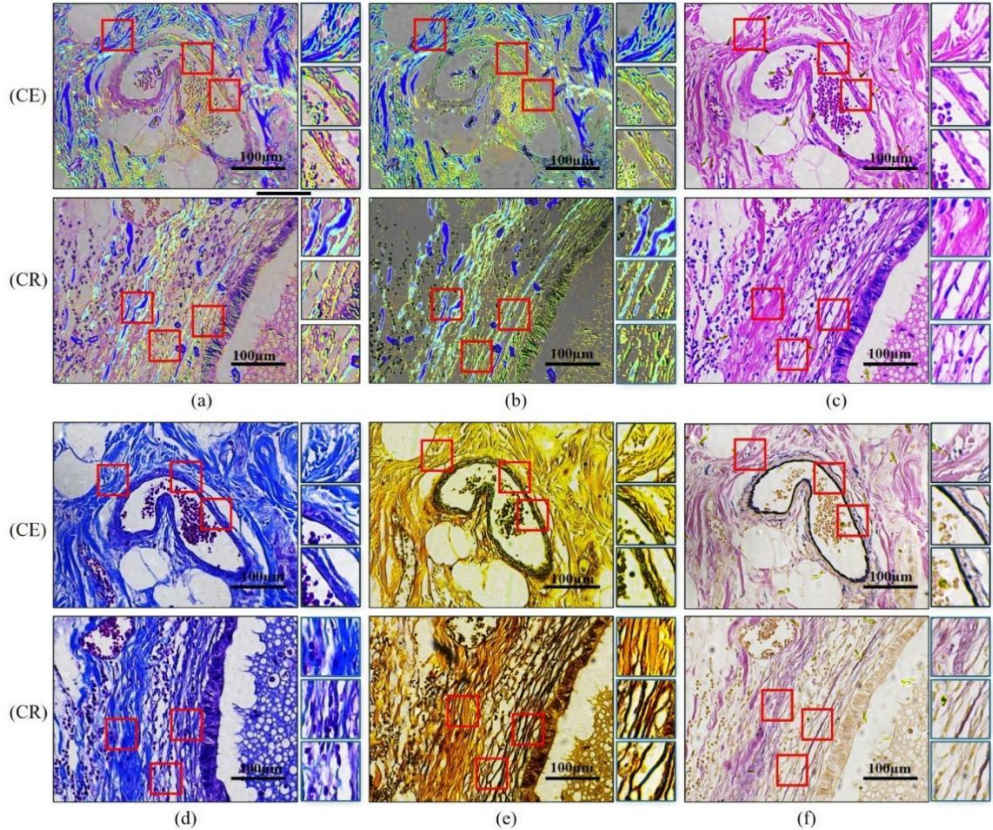

**Fig 4** Pairwise tissue comparison in various staining modalities: (a) Pseudo-stained image fused onto H&E stained image (b) Pseudo-stained image fused onto grayscale representation of the H&E stained image, (c-f) Standard histological stainings for reference: (c) Hematoxylin and Eosin, (d) Masson's Trichrome, (e) Gomori's silver, and (f) Verhoeff-Van Gieson. Left markers indicate pairwise comparison's tissue pair: collagen-elastin (CE) fibers and collagen-reticular (CR) fibers. The red bounding boxes indicate regions that are zoomed in the right-side panels to showcase fine-grained textural details and fiber orientation.

## 4 Discussion

Additionally, to ensure comprehensive analysis and assess the choice of input dimensions for the manifold projection, we evaluated the effect of the ambient dimension by comparing different settings using various statistical distribution metrics. Specifically, we addressed three distinct experimental cases: (1) a smaller set of input parameters composed of a subset of the initial polarization parameters input: MMPD (diattenuation and retardance), MMT (A, b, t) and Mueller matrix elements ($M_{11}$, $M_{14}$, $M_{41}$, $M_{44}$) which are rotation-invariant. This constitutes a low-dimensional scenario where input dimensions and methods are constrained; (2) An extended parameter set including rotation-sensitive parameters derived from Mueller matrix decomposition methods; and (3) a complete parameter set comprising all parameters mentioned. Our previous analysis relied exclusively on rotation-invariant polarization-based parameters, we omitted this aspect when selecting MM elements due to the inseparable nature of the information each of them encode. We excluded clear angle parameters as they do not represent an intrinsic property of tissue structures. We believe that while fast-axis orientation can reveal the nature of a structure in specific application cases, for manifold projection, it may disrupt the inter HDDP distance calculation during the projection process. Parameter selection emphasizes balancing comprehensive multi-angle analysis (incorporating diverse decomposition methods and parameters) with selective parameter optimization, prioritizing highly correlated features to the most dominant property of tissue fibers while discarding less relevant ones to enhance interpretability and accuracy. Balance lies in breadth and precision, capturing enough dimensions while retaining only the most informative subset. To quantitatively assess the performance of each projection, result in differentiating fiber properties, we process the data using the aforementioned procedure to extract statistical measures patch distributions data. Therefore, distributions are compared using statistical metrics including Jensen Shannon divergence (JSD) and Kullback-Leibler divergence[36,37] (KL), Wasserstein distance[38] (WSD) and Kolmogorov Smirnov test[39] (KS) as indicators of separability. These metrics evaluate the divergence between distributions to determine the efficacy of the projection under different input dimensions in distinguishing fibrous tissue classes. Particularly, the value of a projection is heightened, more informative and interesting when its indicator shows a clear separation between distributions. This non-overlapping property is critical, as it provides the basis for accurately distinguishing and classifying different fiber types. Figure 5 recapitulates assessment of the divergence metrics for each input dimension. The layout facilitates comparison as row corresponds to different pairwise tissue category comparisons, and columns represent individual metrics. The plotted values quantify the distributional divergence for each pair (noting that the asymmetric nature of the KL divergence is accounted for by evaluating both KL (Distribution 1 ‖ Distribution 2) and KL (Distribution 2 ‖ Distribution 1). Analysis confirms that projection performance is degraded by perturbations and is not improved by incorporating further redundancy from the originating source. Globally, the original input set exhibits a more favorable distribution divergence, beneficial for the task. In contrast, these additions substantially increased computational cost and running time by more than 20-fold. The expanded memory demand triggered system swapping, causing the process to become I/O-bound rather than compute-bound, while providing no commensurate improvement in projection quality. These findings validate our choice of restricted parameters set in this work. It should be noted that the efficacy of Mueller matrix microscopy is sensitive to sample thickness. In this study, standard histological sections (4 µm) were utilized to minimize multiple scattering and depolarization artifacts that dominate in thicker tissues. However, microscale heterogeneities are present even in thin cross-sections fibrous tissues, manifesting as localized structural stagnation (e.g. uneven density) or irregular H&E dye retention. These variations cause polarization-based parameters to fluctuate. Since the most prominent feature for strongly linear birefringent structures like connective tissue fibers remains the phase delay property, a function of optical path length, the resulting manifold projection can be in first approach approximated as a global retardance map.

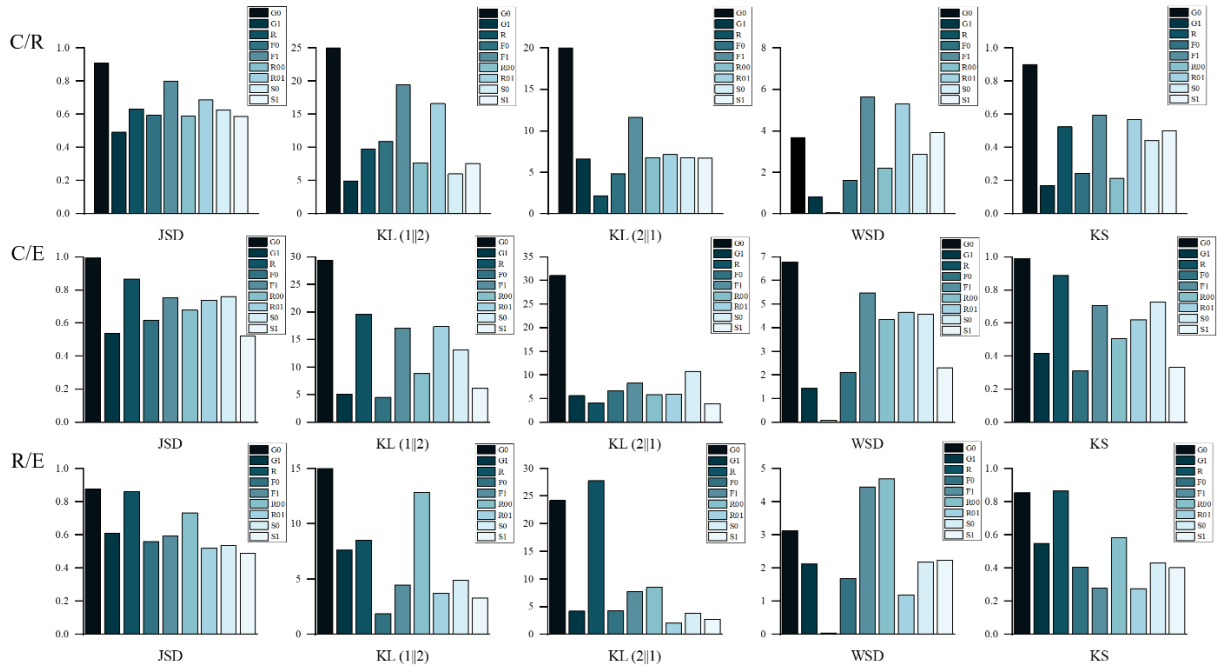

**Fig 5** Column plot comparing distributional divergence metrics for pairwise tissue category analyses. Pairwise tissue comparisons are presented in separate row: C/R ― Collagen/Reticular, C/E ― Collagen/Elastic and R/E ― Reticular/Elastic with the corresponding numerical results of each metrics shown in the columns.

This study serves as a foundational proof-of-concept, validating the efficacy of the restricted parameter set in characterizing tissue micro-architecture. While the initial correspondence between optical signals and fiber structures has been successfully established, it is important to recognize that the current sample size may not yet capture the full spectrum of inter-patient variability inherent in biological tissues. Consequently, future large-scale clinical trials with diverse cohorts will be essential to generalize these findings. Such studies will facilitate the standardization of the color-coding system, ensuring that specific optical signatures can be reliably interpreted as distinct pathological markers, thereby solidifying the method's utility for routine clinical application. Furthermore, the proposed wide-field Mueller matrix approach occupies a distinct operational niche compared to nonlinear modalities like SHG and TPEF. While SHG excels in subcellular specificity via point-scanning femtosecond lasers, its application in high-throughput screening is often constrained by cost and acquisition time. In contrast, our method prioritizes accessibility and scalability. Although the current mechanical modulation imposes limits on temporal resolution, the capability for parallel pixel acquisition over centimeter-scale areas offers a significant advantage for digital pathology, providing a cost-effective alternative for whole-slide quantitative characterization without the need for raster scanning. Therefore, we envision a workflow where Mueller matrix microscopy serves as a rapid primary screening tool, potentially guiding subsequent high-resolution inspections using nonlinear modalities.

# 5 Conclusion

In this work, we utilized Mueller matrix-derived parameters to clarify the organization of extracellular matrix fibers, with a focus on three critical types: collagen (imparting tensile strength in tendons), reticular fibers (offering supportive networks in lymph nodes), and elastic fibers (ensuring vascular flexibility). These fibers are vital for metabolic processes and hold significant promises for direct clinical application. Thus, we employed the nonlinear dimensionality reduction technique UMAP to extract properties from a constructed high dimensional data map, which help cross beyond ideal-case assumptions for each decomposition method and limitation in individual parameter description of sustaining life functions tissues nature. Subsequent statistical analysis revealed distinct behavioral patterns on the projection result for all three types of fiber. Therefore, based on mean patches retrieved from quantitative analysis, we processed pseudo-staining to visualize each of the fibers in different coloration. Current approach in this study differentiates tissue structures through pseudo-staining based primarily on the statistical analysis of pixel mean intensities. This establishes a baseline for distinction as the clustering capabilities of the resulting manifold projection are defined by its fiber-specific intensity peaks, which segregate pixel distributions into discrete feature-space regions corresponding to collagen, reticular, and elastic fibers. These unique regions serve as statistical fingerprints and can be further enhanced by more polarimetry-based robust techniques and advanced modeling to address the remaining challenges in automated segmentation and analysis.

**Disclosures**

The authors declare that there are no financial interests, commercial affiliations, or other potential conflicts of interest that could have influenced the objectivity of this research or the writing of this paper.

**Code and Data Availability**

The code and data that support the findings of this study are available from the corresponding author upon reasonable request.

**Acknowledgements**

National Natural Science Foundation of China (NSFC) Grant No. 62335007.

14. Y. Dong, S. Liu, Y. Shen, H. He, and H. Ma, "Probing variations of fibrous structures during the development of breast ductal carcinoma tissues via mueller matrix imaging," *Biomed. Opt. Express* **11**(9), 4960–4975 (2020). [https://doi.org/10.1364/BOE.397441]
15. He, C., Chen, B., Song, Z. *et al.* "A reconfigurable arbitrary retarder array as complex structured matter". *Nature Communications,* **16**(4902), (2025).[https://doi.org/10.1038/s41467-025-59846-4]
16. Yifei Ma et al. "Using optical skyrmions to assess vectorial adaptive optics capabilities in the presence of complex aberrations". *Science Advances*, **11**(40), eadv7904 (2025). [https://doi.org/10.1126/sciadv.adv7904]
17. H. Zhai, Y. Sun, H. He, B. Chen, C. He, Y. Wang, and H. Ma, "Distinguishing tissue structures via polarization staining images based on different combinations of mueller matrix polar decomposition parameters," *Opt. Lasers Eng.* **152**, 106955 (2022). [https://doi.org/10.1016/j.optlaseng.2022.106955]
18. C. Roa, V. N. D. Le, M. Mahendroo, I. Saytashev, and J. C. Ramella Roman, "Auto-detection of cervical collagen and elastin in mueller matrix polarimetry microscopic images using k-nn and semantic segmentation classification," *Biomed. Opt. Express* **12**(4), 2236–2249 (2021). [https://doi.org/10.1364/BOE.420079]
19. M. Li, S. Cao, and R.-H. Xu, "Global trends and epidemiological shifts in gastrointestinal cancers: insights from the past four decades," *Cancer Commun.*, (2025).
20. P. Danpanichkul, K. Suparan, P. Tothanarungroj, D. Dejvajara, K. Raekwon, Y. Pang, R. Barba, J. Thongpiya, M. B. Fallon, D. Harnois et al., "Epidemiology of gastrointestinal cancers: a systematic analysis from the global burden of disease study 2021," *Gut* **74**(1), 26-34 (2025). [https://doi.org/10.1136/gutjnl-2024-333227]
21. S.-Y. Lu and R. A. Chipman, "Interpretation of mueller matrices based on polar decomposition," *J. Opt. Soc. Am. A* **13**(5), 1106–1113 (1996). [https://doi.org/10.1364/JOSAA.13.001106]
22. J. J. Gil, R. Ossikovski, and J. J. Gil, *Polarized light and the Mueller matrix approach*. CRC press (2022).
23. H. He, N. Zeng, E. Du, Y. Guo, D. Li, R. Liao, and H. Ma, "A possible quantitative mueller matrix transformation technique for anisotropic scattering media/eine mogliche quantitative muller-matrix-transformations-technik fur anisotrope streuende medien," *Photonics Lasers Med.* **2(2)**, 129–137 (2013). [https://doi.org/10.1515/plm-2012-0052]
24. S. R. Cloude, "Group theory and polarisation algebra," *Optik (Stuttgart)* **75**(1), 26–36 (1986).
20